\documentclass[sigconf,nonacm]{acmart}
\usepackage{algorithm}
\usepackage{algorithmic}
\usepackage{amsmath}
\usepackage{balance}
\usepackage{cleveref}
\usepackage{enumitem}
\usepackage{fancyvrb}
\usepackage{float}
\usepackage{graphicx}
\usepackage{hyperref}
\usepackage{listings}
\usepackage{mathtools}
\usepackage{microtype}
\usepackage{multirow}
\usepackage{subcaption}
\usepackage{xcolor}
\usepackage{xspace}

\newtheorem{example}{Example}
\newtheorem{definition}{Definition}

\newlength{\listingindent}                %
\definecolor{applegreen}{rgb}{0.55, 0.71, 0.0}
\definecolor{asparagus}{rgb}{0.53, 0.75, 0.48}
\definecolor{deeplilac}{rgb}{0.6, 0.33, 0.73}
\definecolor{maroon}{rgb}{0.5, 0.0, 0.0}
\definecolor{royalfuchsia}{rgb}{0.79, 0.25, 0.37}
\definecolor{maroon}{rgb}{0.69, 0.19, 0.38}
\definecolor{grey}{rgb}{.6, ,.6, .6}

\newcommand{\eat}[1]{}

\newcommand{\stitle}[1]{\smallskip\noindent\textbf{#1}}

\newcommand{\red}[1]{\textcolor{red}{#1}}
\newcommand{\blue}[1]{\textcolor{blue}{#1}}
\newcommand{\purple}[1]{\textcolor{purple}{#1}}
\newcommand{\grey}[1]{\textcolor{grey}{#1}}

\newcommand{\codesize}{\fontsize{7}{8}}
\newcommand{\code}[1]{{\codesize\texttt{#1}}\xspace}

\newcommand{\pisys}{\textsc{PI2}\xspace}
\newcommand{\pisysfull}{\textsc{Precision Interfaces 2}\xspace}
\newcommand{\pvdd}{\textsc{Physical Visualization Design}\xspace}
\newcommand{\pvd}{\textsc{PVD}\xspace}

\newcommand{\dig}{\textsc{DIG}\xspace}
\newcommand{\digg}{\textit{Data Interface Grammar}\xspace}

\AtBeginDocument{%
  }

\begin{document}

\title{DIG: The Data Interface Grammar }

\author{Yiru Chen}
\affiliation{%
  \institution{Columbia University}
  \streetaddress{1 Th{\o}rv{\"a}ld Circle}
  \city{New York}
  \state{NY} 
  \country{USA}}
\email{yiru.chen@columbia.edu}

\author{Jeffrey Tao}
\affiliation{%
  \institution{Columbia University}
  \city{New York}
  \state{NY} 
  \country{USA}}
\email{jat2164@columbia.edu}

\author{Eugene Wu}
\affiliation{%
  \institution{Columbia University}
  \city{New York}
  \state{NY} 
  \country{USA}}
\email{ewu@cs.columbia.edu}

\renewcommand{\shortauthors}{Yiru Chen et al.}

\begin{abstract}
Building interactive data interfaces is hard because 
the design of an interface depends on the data processing needs for the underlying analysis task,
yet we do not have a good representation for analysis tasks.
To fill this gap, this paper advocates for a {\it Data Interface Grammar} (\dig) as an intermediate representation of analysis tasks.
We show that \dig
is compatible with existing data engineering practices, compact to represent any analysis,
simple to translate into an interface design,
and amenable to offline analysis.
We further illustrate the potential benefits of this abstraction, such as automatic interface generation, 
automatic interface backend optimization, tutorial generation, and workload generation. 
\end{abstract}

  

\maketitle

\section{Introduction}
\label{sec:intro}

Interactive data interfaces are essential for data exploration and analysis~\cite{facets, icheck, Murray2013TableauYD,Kandel2011WranglerIV, chen2022tsexplain}. However, designing a new interface is a multi-step process that includes determining the queries that are appropriate for the analysis task, as well as the parts of the queries that users should be able to change. Once the analysis has been determined, the interface designer can now choose the appropriate visualizations, interactions, and layouts to design the interface. These two steps are closely related and need to be kept in sync, yet require distinctly specialized skill sets: to write complex queries over complex data sources, and to design and implement a usable and effective interface.  Further expertise in system optimization is needed to ensure that the resulting interface is responsive as the data grows.

Is there an intermediate abstraction of an interface's underlying data needs that can decouple these tasks, so that data practitioners can focus on expressing complex analysis tasks, visualization designers can focus on interface design, and backend engineers can focus on optimization?

\begin{figure}[t]
      \centering
      \includegraphics[width=\columnwidth]{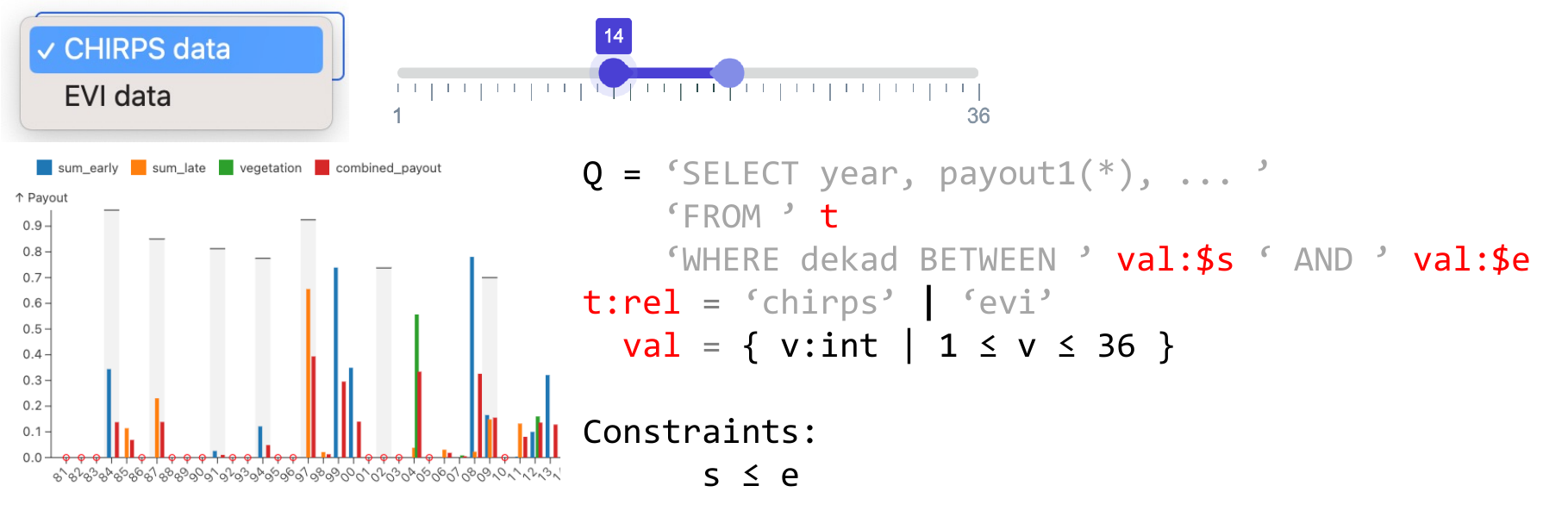}
      \caption{Subset of the Open Policy Kit interface to design drought insurance policies for rural farmers~\cite{greatrex2015scaling}.   The user can choose from Chirps and EVI rainfall data sources, tune the start of the measurement period, and see how different payout calculations aligns with historical droughts and their own expectations.  The dropdown changes the underlying query's \texttt{FROM} clause, and the sliders change a range filter condition.  Each interaction issues a new query to the database, whose result is rendered in the chart.   The \dig code concisely describes the interface's data needs.  }
      \vspace{-.2in}
      \label{fig:intro}
  \end{figure}

Let us first examine the different ways a developer might build an interface today. \Cref{fig:intro} is a simplified subset of a drought insurance design tool used to protect rural farmers~\cite{greatrex2015scaling}.  The simplest approach is to predefine all possible queries in the application ahead of time, and when the user interacts with the interface, we identify which query to execute.   Although the queries can be optimized ahead of time, this requires enumerating a combinatorial number of possible queries (e.g., 1332 = 2 for the dropdown * 666 for the slider).   Parameterized queries allow literals in the WHERE clause to be wildcards.  This compactly expresses the slider interaction and can be optimized offline~\cite{ioannidis1997parametric}, but cannot express arbitrary structure changes in the query (e.g., the dropdown).   
In short, there exist tools to create {\it and optimize} very simple data interfaces where the interactions largely correspond to filters or where the user simply cannot express very much.    
Beyond this, the developer must resort to constructing query strings in the application, which is highly flexible but not amenable to analysis. 

What criteria should an intermediate abstraction satisfy?  We believe that it should (C1) compactly represent any analysis task that a developer may wish to express, (C2) have a well-defined correspondence to interactive interfaces composed of charts, widgets, and interactions,  and (C3)  be amenable to offline analysis for e.g., optimization, interface synthesis.

It is easy to see that existing approaches do not satisfy these criteria.   Predefining every query is neither compact nor expressive, parameterized queries are compact but only express simple query transformations, and programmatically constructing SQL is expressive but not analyzable.  Other works on interactive data interface benchmarks~\cite{Eichmann2018IDEBenchAB,battle2020database} model the interface in order to generate query workloads but are limited to SPJA queries and cross-filter interactions. Business Intelligence(BI) and visualization tools (e.g., Tableau~\cite{tableau}, Power BI~\cite{Powerbi}, Vega-lite~\cite{Satyanarayan2016ReactiveVA, Satyanarayan2018VegaLiteAG}) are primarily focused on data cube-like operations.   

In this paper, we examine two observations.
First, the queries that an interface expresses can be compactly represented as a grammar.  A grammar is a set of production rules that define a valid program; each production rule defines a set of choices that encode the allowable program variations. That grammar may be a single production rule that chooses from a small enumeration of predefined queries, the entire language (e.g., SQL), or a language subset specific to an analysis.   
Second, the design of a data interface has a direct correspondence to the grammar: interactions make choices in the grammar, and when all choices in the grammar have been made, the grammar is equivalent to a syntactically valid query string that the database executes.  In other words, interactions navigate the space of syntactically valid queries expressible by the interface.  

To this end, we propose \dig, a {\it Data Interface Grammar} that extends Parsing Expression Grammars (PEG) with annotations specific to data programs.
For instance, the \dig program for \Cref{fig:intro} is a query string (\grey{gray text}) where nonterminals encode program variations: \code{t} chooses the relation name, and \code{val} chooses an integer between 1 and 36.  \code{t} is annotated to be a relation name, and \code{s}$\le$\code{e}.   In the interface, these choices are respectively bound by the dropdown and range slider.

\dig satisfies our desired criteria.  
Since it extends a formal grammar, it compactly express any set of queries useful for a task (C1), and defines a direct correspondence to interactive interface designs (C2).  %
Finally, since \dig encodes the the entire space of possible programs, it is amenable to offline analysis, and \Cref{sec:problem} outlines examples such as interface synthesis and physical optimization(C3).

In the rest of this paper, we will first introduce {\it Data Interface Grammar} (\dig)  and illustrate its correspondence with interactive interfaces.   We further comment on its connections with existing data pipeline and analysis representations (\Cref{sec:dig}).  
We then describe how \dig simplifies interface creation via real-world examples (\Cref{sec:vision}), and finally highlight the benefits of the \dig representation for solving a number of challenging data interface problems (\Cref{sec:problem}).

\section{Data Interface Grammar}
\label{sec:dig}

A data interface helps the user navigate a space of useful data programs (e.g., SQL) through interactive controls.  This section first presents \dig, a Data Interface Grammar, to express this set of data programs in a simple, analyzable manner, and then defines the set of valid interfaces that express a given \dig program.  These definitions form the basis for useful applications like interface synthesis, physical visualization optimization that we describe in \Cref{sec:problem}.

\subsection{\dig Definition} 

\dig is a \digg that defines the syntactic structure of queries 
that an interface wishes to express: the set of queries parsable by the grammar.  
Given that existing data query languages such as SQL, PRQL~\cite{prql}, and Pandas have well-established grammar definitions, \dig is a superset of the widely-used Parsing Expression Grammar (PEG). By extending PEG, we both build on decades of research and tooling and simplify the ability to port existing PEG-based languages to \dig.

\noindent We formally define  \dig = $\{N, \Sigma, P, e_S, C\}$ as follows, where the sub-grammar rooted at each starting rule parses a set of queries:
\begin{itemize}[leftmargin=*]
    \item a finite set of nonterminals $N$;
    \item a finite set of terminals $\Sigma$ that is disjoint from $N$;
    \item a finite set of parsing rules $P$;
    \item a finite set of starting rules $e_S$, each not referenced by any other rule;
    \item a set of constraints C.
 \end{itemize}

\stitle{Terminals.}
Similar to typical grammars, \dig matches terminals to valid strings expressible by regular expressions.  Although regular expressions are useful for matching string literals, most interactions (e.g., sliders, dropdowns, visualization selections) are typed and limited to a domain of valid values that regular expressions cannot distinguish.   Thus, \dig also supports \textit{domain terminals} that may reference the underlying database.    
\begin{itemize}[leftmargin=*]
\item Predicate Domain: \code{A = \{var:type | <predicate>\}}.
\item Query Domain: \code{A = \{SELECT QUERY\}}.
\end{itemize}
\noindent A predicate domain specifies a typed variable along with a boolean expression that must evaluate to true for a value to be valid.  For instance, \code{val = \{x:int | x$\in$[1,36]\}} specifies the terminal as an integer between \code{1} and \code{36}.  Note that a regular expression pattern \code{p} is expressible as a predicate domain \code{\{s:str | s matches p\}}.

A query domain specifies a query over the database; the terminal must be an element in the query result.   For instance, \code{prods = \{SELECT name FROM products\}} ensures that the terminal is a valid product name.   The data types may be structured as well, for instance {\code{X = \{ SELECT fname, lname FROM users \} }} would choose from the first and last names of existing users.   This formulation serves as hints for the interface to choose a good interaction for the rule, and as input validation rules to guarantee syntactically correct programs.   

Following second order languages like SchemaSQL~\cite{lakshmanan1996schemasql}, we additionally support special string types to express relation names (\code{rel}) and attribute of a relation (\code{attr[str:rel]}) where it is optionally parameterized by a relation name.   Thus the following restricts \code{name}s to attribute names in two relations:
{\small\begin{verbatim}
 sources = { s:rel | s in ['usproducts', 'euproducts']}
    name = { s:attr[sources] }
\end{verbatim}}
 
\stitle{Rules.}
Each rule in $P$ is structured as $A = e$, where $A$ is a non-terminal and $e$ is a parsing expression composed of a reference to a non-terminal, a terminal (e.g., a string literal), or an expression composed of either a sequence $e_1 e_2$, selection $e_1 | ... | e_n$, or zero-or-more $e^*$ operator\footnote{PEG operators like \code{AND} and \code{NOT} can be omitted since \dig is not used for parsing.}.  Selection implicitly has a domain $[1,n]$ that specifies which subexpression is selected, and zero-or-more's domain is the natural numbers, which specifies the number of repetitions. 
Other patterns, such as $e^+$ and $e?$, are reducible to these operators.

A non-terminal $A$ on the left side of a rule can optionally be typed by adding the suffix \code{:type}.  For instance, \code{t:rel} in \Cref{fig:intro} specifies that \code{`chirps` and `evi`} are relation names.   Type violations result in a parsing error.

\stitle{Naming.}
Naming is necessary for defining constraints and interface mappings next, thus we now introduce annotations and choice variables. 
An annotation assigns a variable name to a non-terminal reference by appending \code{:\$varname} to the reference.  For instance, \Cref{fig:intro} assigns the two \code{val} references to \code{s} and \code{e}.  If a reference is not annotated, \dig assigns a unique name by appending a unique number to the non-terminal name (e.g., \code{val1}).  

Unfortunately, variables alone are not sufficient because the same non-terminal can be referenced multiple times.   Consider the following rules:

{\small\begin{verbatim}     A = B:$v1 B:$v2        B = C:$v3          C = \d+\end{verbatim}}
\noindent The variable \code{v3} is ambiguous because both \code{v1} and \code{v2} reference it.    Thus, we define a variable's fully qualified name as the path from the root of the \dig to the variable, where each element in the path is a non-terminal reference. 

All variability in \dig is expressed by non-terminals that expand to a predicate or query domain, selection expression, or zero-or-more expression.  We use the term {\it choice variable} to refer to the fully qualified reference to such a non-terminal.  
For instance, \code{v1}, \code{v2}, \code{v1/v3}, \code{v2/v3} are the choice variables in the above example.   Further, let $D_c$ be the domain of a given choice variable, as defined in the \textit{Terminals} and \textit{Rules} paragraphs above.

\stitle{Constraints.}
The developer can specify boolean expressions over choice variables; these constraints are evaluated when the user performs an interaction to determine validity.     For instance \code{s}$\le${e} in \Cref{fig:intro} ensures that start should be less than or equal to the end of the range.   Two terms assigned to the same variable name implies an equality constraint.     \dig handles equality constraints between variables (e.g., \code{s = e}) in a special way: if one variable is bound to a value $v$, then the other is updated to $v$ as well; if both are updated then we check if they are equal.

\subsection{Valid Interfaces for a DIG}
An interface renders query results and lets the user navigate the space of valid queries.
Since choice variables encapsulate all variability, a valid interface is one whose interactions can bind values to the set of choice variables.   Once they are all bound, the grammar reduces to an executable query string, and the interface renders its evaluation result(s).   We will first define interactions and how they {\it cover} choice variables, and then define the set of valid interfaces for a \dig grammar.   In practice, each starting rule in a grammar represents a separate query, and the interface will render the results of each query; this extension is straightforward and we assume a single root for clarity.

\stitle{Interfaces and Interactions.}
An interface $UI=(V, I)$ consists of a view $V$ (e.g., a table, a visualization, a paragraph) that renders the output of the starting rule and a set of interactions $I$.  We model an interaction $i = (T_i, D_i)\in I$ by the state it can express. $T_i$ is its type (e.g., dropdown, slider) and $D_i(a_1,...)$ is its domain with schema $(a_1,...)$. For instance, the domain for a dropdown with $n$ options is $[0,n]$; for a text box is the set of all strings (perhaps up to a specified length); for a slider is the set of numbers between the min and max; and for a 2-D brush interaction in a scatter plot is the set of bounding boxes in the chart.  An interaction's developer is responsible for defining its domain.  Note that our interface model supports arbitrary layout because layout does not affect the interface's expressiveness\footnote{Chart layout (e.g., faceting/small multiples) may affect the set of interactions the chart can express, but is encapsulated by the chart.}

Let a mapping $M_{i,c} = \{a_i\to a_c | a_i\in schema(D_i)\land a_c\in schema(D_c) \}$ map attributes in the interaction's domain to attributes in the choice variable's domain, and the mapping's projection $\pi_M(D_i)$ be the subset of attributes in the interaction's domain that have a mapping.  An interaction $i$ is said to {\it cover} a choice variable $c$ if 1) every attribute in $D_c$ is mapped to in $M_{i,c}$, and 2) the interaction's domain is a superset of the choice variable's domain: $\pi_D(D_i)\supseteq D_c$.   These ensure that all possible assignments to $c$ can be expressed in the interface.   Given these definitions, we are now ready to define a interface validity.  

\begin{definition}[Valid Interface]
An interaction $UI$ is valid for a \dig grammar $G$ if every choice variable in $G$ is covered by at least of interaction in $UI$, and every root rule is rendered by at least one view.
\end{definition}

\begin{example}
  \Cref{fig:intro} contains two interactions and two mappings. 
  The dropdown maps its selected index to the choice variable \code{t}; since the dropdown is initialized with set of choices in \code{t} (e.g., ``chirps'', ``evi''), their domains will be identical - $[1,2]$.
  The range slider maps the left slider handle to \code{s} and the right slider handle to \code{e}; the slider's domain is $\{(l,r) | l\in[1,36]\land r\in[1,36]\}$, which matches the predicate domain and constraints over \code{s} and \code{e}.
\end{example}

\begin{figure}[h]
    \centering
    \includegraphics[width=\columnwidth]{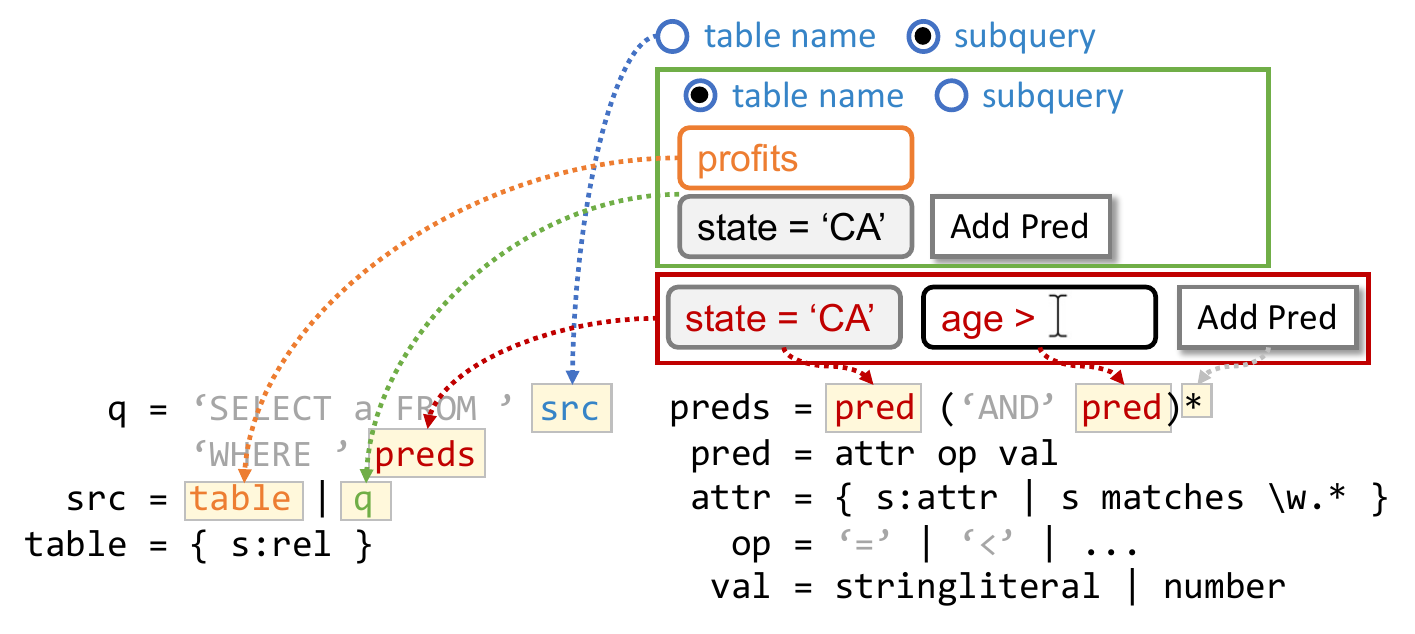}
    \vspace{-.1in}
    \caption{A valid interface for a \dig grammar with text inputs and recursion.  The \code{pred} rule parses each predicate text box to bind \code{attr}, \code{op}, and \code{val}.   The radio button chooses between a base table or subquery as the source, and choosing the subquery instantiates a recursive copy of the interface. }
    \vspace{-.2in}
    \label{fig:recursion}
\end{figure}

\stitle{Text Inputs and Parsing.}
Text inputs are a special type of interaction because they can, in principle, produce arbitrary strings that are interpreted as query substrings rather than string literals.  For instance, \Cref{fig:recursion} is a query builder where the user types in predicate expressions, and clicks on ``add pred'' to add additional conjunctive clauses.   The text input is parsed by the \code{pred} rule, which implicitly binds the attribute, operator, and value.  

For these reasons, a text input\footnote{In general, this can be any interaction whose domain is all strings. } can map to any term $t$ in a \dig grammar.   Any input string will first be parsed and validated by the subgrammar rooted at $t$.  The parsing process implicitly binds all of the choice variables in the subgrammar, and all parsing errors or constraint violations are passed to the interaction in order to surface as error messages.   

This functionality is helpful for several reasons.  First, \dig can automatically perform parsing and validation such that any text input is guaranteed to be syntactically correct and naturally prevents issues such as SQL injection.   Second, every \dig statement is guaranteed at least one valid interface: one where a text input maps to the root of the grammar, which is equivalent to a typical console-based interface.   Third, it enables a progressive interface design process, where starting from the default text-based interface, more specialized interactions are added to the interface to ``carve out'' more and more choice variables.

\stitle{Recursive Rules.}
So far, we have implicitly assumed that the \dig grammar is hierarchical and non-recursive.  However, \dig allows recursion in the rule set.  For instance, SQL allows nested queries anywhere a value or relation is expected.    How does recursion map to a valid interface?  We outline three categories of approaches.

The first approach is to simply map the first external reference of the recursive rule set to a text input.  However, this may reduce the interface to a single text box.  The second approach is to enforce a maximum recursion depth and ``unroll'' the recursion; this effectively produces a new grammar without this recursion that we can map to a valid interface. 
The third approach is inspired by existing query builder interfaces that support nested queries: they typically instantiate a ``new'' query builder interface to represent the nested query.  In \dig, we first ignore the expression $e_r$ that introduces the recursion and determine a valid interface $UI$ for the remaining rules in the recursive rule set.  We then map a button to $e_r$ that, when clicked, instantiates a new instance of $UI$ in the interface.     

\Cref{fig:recursion} illustrates recursion in the \code{src} rule, which specifies a table name or a subquery with the same structure.   The interface lets the user choose between the options using a radio button.   If the user chooses table name, they fill a text box with the relation name (e.g., ``profits''), which is validated by the \code{table} rule.  Otherwise, we instantiate a nested interface containing the radio buttons and set of predicates.   

\subsection{Cross-filter Example}

\begin{figure}
    \centering
    \includegraphics[width=0.8\linewidth]{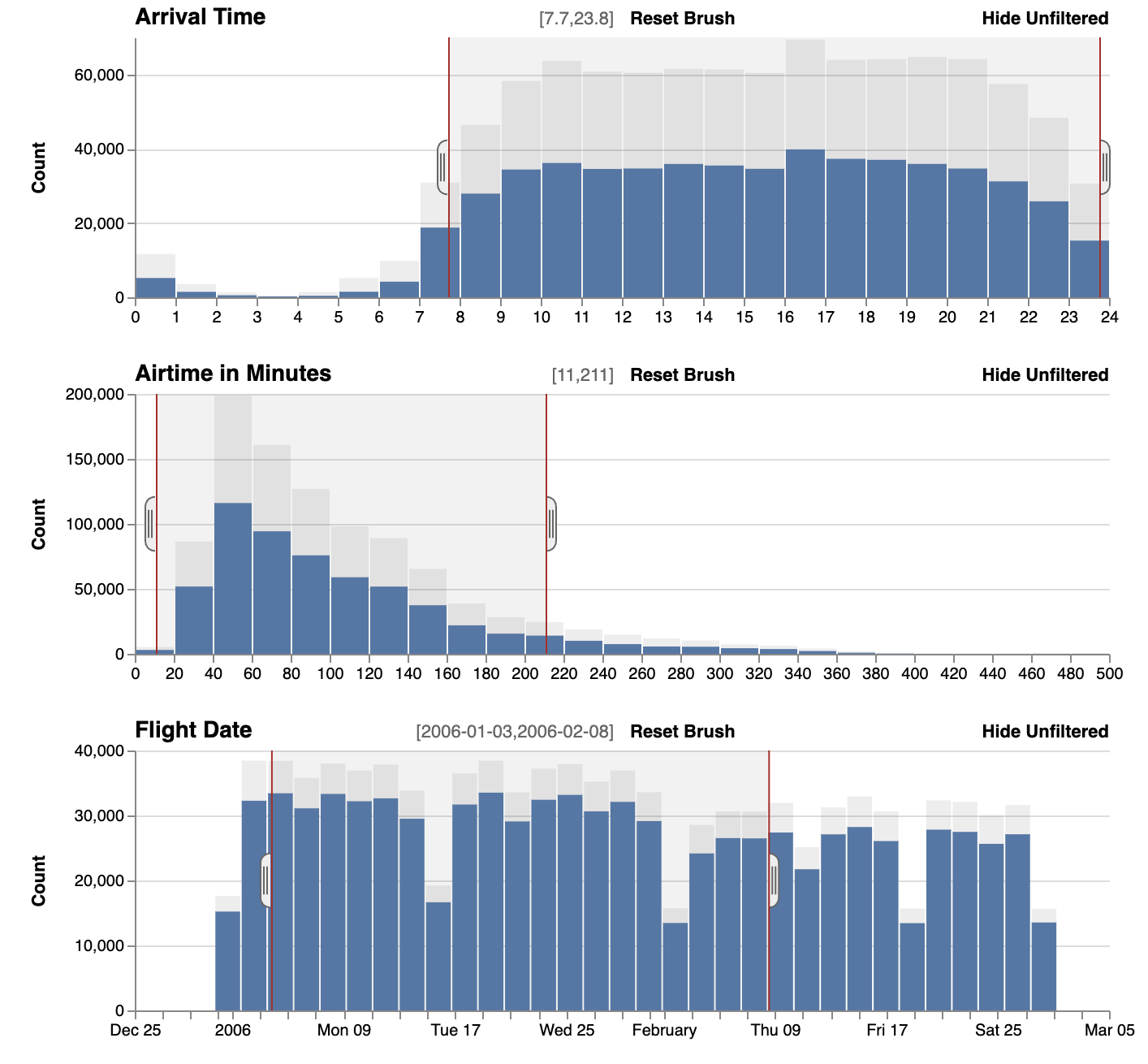}
    \caption{Cross-filter renders histograms over different dimensions. Brushing over a chart adds a filter to the other charts based on the selected range.  }
    \label{fig:falcon}
\end{figure}

We now use the popular cross-filter visualization~\cite{crossfilterjs} to illustrate \dig end-to-end on a real-world example.  Each bar chart in cross-filter renders an aggregation grouped on one attribute in the underlying dataset.  For instance,  \Cref{fig:falcon} shows three bar charts grouped on arrival time, airtime, and date, respectively.  Brushing over a chart grouped on \code{attr} adds a predicate that filters over \code{attr} to the other charts; the filtered aggregates are rendered as an overlay, while the unfiltered results are gray in the background.

The following \dig grammar describes the rules to render the arrival (\code{q1}) and airtime (\code{q2}) charts; it omits constraints and the rules for the date chart.   
\begin{Verbatim}[fontsize=\footnotesize,commandchars=\\\{\}]
    q1_bg = \grey{'SELECT arrival, count() FROM flights GROUP BY arrival '} 
    q2_bg = \grey{'SELECT airtime, count() FROM flights GROUP BY airtime '} 
        ...
       q1 = \grey{'SELECT arrival, count() FROM flights WHERE '} 
            \red{p_airtime:$pair} \grey{' AND '} \blue{p_date:$pd } \grey{'GROUP BY arrival'}
       q2 = \grey{'SELECT airtime, count() FROM flights WHERE '} 
            \purple{p_arrival:$parr} \grey{' AND '} \blue{p_date:$pd } \grey{'GROUP BY airtime'}
        ...             
\purple{p_arrival} = true | \grey{'arrival BETWEEN '} \purple{arr}:$arrs \grey{' AND '} \purple{arr}:$arre
\red{p_airtime} = true | \grey{'airtime BETWEEN '} \red{air}:$airs \grey{' AND '} \red{air}:$aire
   \blue{p_date} = true | \grey{'date BETWEEN '} \blue{date}:$s \grey{' AND '} \blue{date}:$e
      \purple{arr} = \{ SELECT arrival FROM flights \}
      \red{air} = \{ SELECT airtime FROM flights \}
     \blue{date} = \{ SELECT date FROM flights \}
\end{Verbatim}
The \code{\_bg} starting rules define the background unfiltered results, while \code{q1} and \code{q2} define the overlay filtered queries.   Each query is filtered by a conjunction of predicate rules (e.g., \code{p\_airtime}); and each predicate such as \code{p\_airtime} either evaluates to \code{true}, meaning the airtime chart is not brushed, or a \code{BETWEEN} clause, meaning that the airtime chart is brushed and the start and end of the brush range map to \code{airs} and \code{aire}.   Notice that \code{p\_date} in \code{q1} and \code{q2} are both named \code{pd} to ensure that their bindings are identical.

\subsection{Tool Compatibility}
A benefit of \dig is that it is compatible with existing data engineering practices.  For instance, data pipelines and analyses are increasingly expressed as a DAG of SQL views using tools like DBT~\cite{dbt}.     This is useful because data engineers can define these DAGs, while business analysts and data consumers can use these views in visualization and data science tools.  
Each  DAG node is called a {\it model} and expressed as a Jinja template that, when evaluated, returns a SQL string.  The template can reference custom variables and call logic to change the query by assigning values to the variables in a configuration file.    

For instance, the following model uses the variables \code{region} to choose the input table (specified using the \code{ref()} function), and \code{age} to change the filter.  \code{region} may be set to \code{USA} or \code{EUR}, themselves are names of other models.
\begin{Verbatim}[fontsize=\footnotesize]
    SELECT cty, sum(profit) FROM {{ref(var("region"))}}
    WHERE age > {{var("age")}}    
\end{Verbatim}
DBT models that \code{ref()}, variable, and branching logic can automatically translate into \dig grammars. \code{ref()} translates into non-terminal reference to either a base relation/view or the starting rules for grammars translated from other models; a variable translates into a terminal rule; and branching logic translates into a selection rule (e.g., \code{e1 | e2 | ..}) with one option for each branch; if the condition expression references a variable, we evaluate the expression dynamically to decide which branch to choose.  

For instance, the above model translates into the following, where we assume \code{usa} and \code{eur} are the starting rules for their respective DBT models.
\begin{Verbatim}[fontsize=\footnotesize,commandchars=\\\{\}]
       q = \grey{'SELECT cty, sum(profit) FROM '} \red{t} \grey{' WHERE age > '} \blue{age}
       \red{t} = usa | eur          usa = ...           eur = ...
     \blue{age} = \{ n:int | n > 0 \}

\end{Verbatim}

\section{Vision: \dig-based Interface Creation}
\label{sec:vision}

So far, we have described \dig as a compact and expressive abstraction that naturally maps to interactive interfaces.
How can such an abstraction change how we design, implement, and use new data-oriented interfaces?
Here, we sketch a potential development cycle that \dig can enable.  
The next section sketches our progress towards this vision.

\stitle{Design.}  Barb wants to create a new data interface to analyze user signup flows, and decides to use \dig.  One option is to manually write a \dig grammar.  Alternatively, she might induce a grammar from existing user signup analyses by extracting queries from e.g., Jupyter notebooks, DBMS query logs, or other database-backed applications, or by translating a natural language description of her analysis goals into a \dig grammar.

Her design tool then automatically synthesizes a custom interactive interface.  She likes the overall design, but resizes the canvas to fit a smaller screen, and specifies that the interface should be more expressive.   The synthesized interface updates, and she re-positions the charts and widgets to fine-tune the layout.  

\stitle{Implementation.}  Barb now connects the design tool to the the user signups database.   If her dataset is small, the design tool can load the database into memory and either run the interface, or export to a web application.  However, if the dataset might grow over time or if it resides in a cloud database (which optimizes for throughput rather than query latency), then Barb potentially needs to engineer an entire client-server system.   However, Barb does not have the time, desire, nor expertise to make all of the decisions about which DBMS, data structures, and optimization techniques to employ so that the interface is responsive.

Instead, Barb gives the design tool her budget, and specifies her desired responsiveness for the different interactions.    The tool uses metadata about the underlying database to estimate how much resources are needed to meet her responsiveness goals.  The proposed architecture requires materializing and caching 7GB of data structures~\cite{pahins2016hashedcubes, graefe2011modern} in server-side memory, which costs \$35/month.  Barb thinks it's too expensive, and moves part of the interface related to post-signup actions to a separate page; this relaxes some of the interactivity constraints, and reduces the sizes of the data structures to 2GB and costs to \$15/month.  When she accepts the recommendation, and the design tool allocates a cloud server, instantiates the data structures and execution plans, and hosts an end point for the new interface.

\stitle{Use.}
Barb knows that learning to use the new interface can be hard for users, so she records herself performing some example analyses.  A new user plays with the interface for a bit, gets confused, and then watches a recording.  Half way through, he wonders how he can get to that point without reloading the interface and starting from scratch.  He clicks a ``show me how'' button, and the interface dynamically creates a tutorial from where he currently is to the point in the recording.   After following the tutorial, he asks {\it ``show user flows for only adults above 50''} in natural language; it automatically aligns this with the grammar's structure,  translates the natural language input into the appropriate choice variable bindings, and the interface walks through the interactions needed to perform this request.

\section{Progress So Far}
\label{sec:problem}

\dig introduces novel problems to improve how interfaces are created, optimized, and used.  We now outline for example problems that we have explored in current or prior work.  

\subsection{Automatic Interface generation}
\Cref{sec:dig} defined the set of valid interfaces that can be mapped from a \digg, and enables the potential to automatically explore and generate valid interfaces for a given \dig grammar.
It is also possible to transform the grammar to induce new sets of valid interfaces.    
Consider the following example based on our recent work called \pisysfull(\pisys)~\cite{PI2}:

\begin{example}[Interface Generation]
\Cref{fig:interfacegeneration}(a) is an initial \dig grammar and a corresponding valid interface.   The grammar expresses four queries that each differs in the filter predicate string; the interface simply selects one of the predicate strings using radio buttons.   We can rewrite the grammar to the equivalent grammar in \Cref{fig:interfacegeneration}(b) by factoring out the ``\code{=}'' character from each predicate and creating separate rules for the left and right sides.   The corresponding interface has two sets of radio buttons, one to choose the attribute and one to choose the value.   Although this appears trivially similar, we might now apply {\it generalization} rules to e.g., let \code{var} match any number, or to lift \code{attr} to an attribute type.  These rules increase the expressiveness of the resulting grammar,  and consequently, the set of valid interfaces that a cost model might pick from.  

\begin{figure}
    \centering
    \includegraphics[width=\columnwidth]{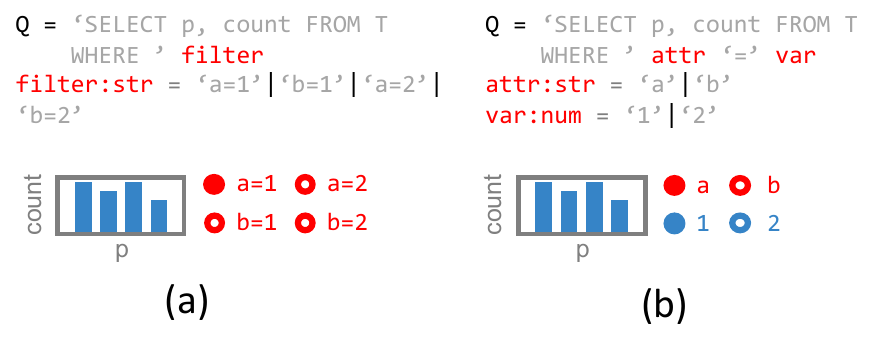}
    \vspace{-.3in}
    \caption{Transforming the \dig grammar changes the set of valid interfaces.  }
    \vspace{-.2in}
    \label{fig:interfacegeneration}
\end{figure}
\end{example}

\stitle{Where Do {\dig}s Come From? } There are many ways to generate a \dig grammar.  In our prior work, we have explored  a sequence of SQL queries from database logs \cite{PI2,chen2020monte}, analyses in notebooks \cite{tao2022demo}, or query models in DBT~\cite{dbt}. Alternatively, it can also be generated from a large language model~\cite{Chen2022NL2INTERFACEIV}, as LLMs are proficient at generating text. For instance, in \Cref{fig:interfacegeneration}, the \dig could be generated from a natural language query such as "How is the total count for different {\it p} when {\it a} is one versus when {\it b} is two?"

\subsection{Automatic Backend Optimization}

Users care about interactivity, and can detect even milliseconds of interaction delay~\cite{jota2013fast}.   As a result, designers must make complex trade-offs between the interface design, levels of responsiveness for different interactions, and the systems and resource implications to guarantee those levels of responsiveness.

Ideally, a designer can label different interactions with their latency constraints and allow an automated tool to check their feasibility and resource requirements.  This is not straightforward today.   Physical database advisors~\cite{agrawal2006autoadmin} take a sample of queries as input, but individual nor sets of queries do not map directly to interactions because, as we have shown, interactions {\it transform} targeted portions of a query.

In contrast, \dig naturally models interactions based on the non-terminals they bind; annotating each interaction with latency expectations is now straightforward.   
This further offers a complete picture of the interface's data processing requirements, as this annotated \dig grammar expresses the universe of possible queries along with their latency requirements.
Given this annotated grammar, we can identify visualization-specific physical data structures to materialize and maintain, along with a placement and query execution plan that spans the client, server, and cloud DBMS, that guarantees these latency requirements.   We term this problem \pvdd (\pvd).

\begin{figure}[h]
        \centering
        \includegraphics[width=0.85\columnwidth]{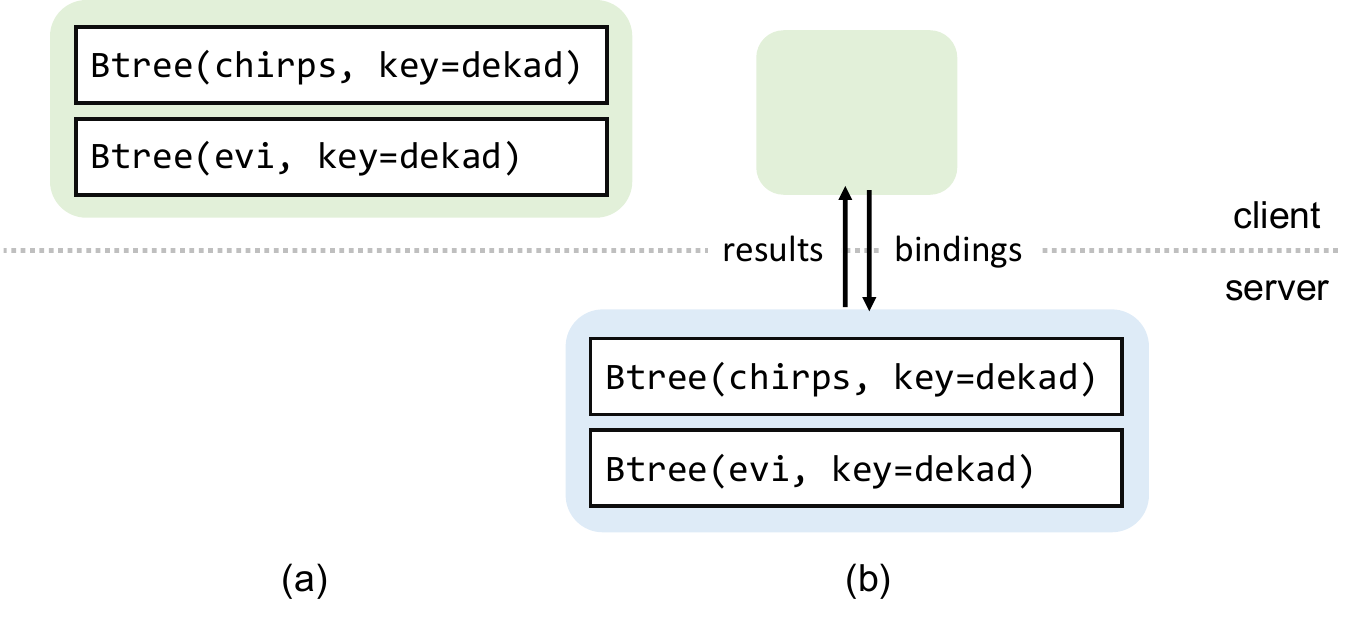}
        \caption{ Two physical designs for the interface in \Cref{fig:intro}. }
        \label{fig:pvd}
        \end{figure}

\begin{example}[Physical Visualization Design]
Consider the interface in \Cref{fig:intro}.   The designer specifies that the slider shold respond in 10ms and dropdown in 100ms, and that the client and server memory constraints are set to 5GB and 50GB, respectively.

\Cref{fig:pvd} shows two potential physical designs.
The first suggestion (a) might be to materialize BTree data structures over the \code{chirps} and \code{evi} datasets on the client in order to execute the slider range interactions as index lookups.  If the estimated data structure sizes are less than 5GB, then this option is desirable.  
If their sizes grow too large, then moving their placement to the server may be preferable (b).  This incurs network communication latencies, but the index lookups may be faster due to a faster server CPU.  

\end{example}

\subsection{Tutorial Generation}
When encountering a new interface, the user must both learn how the interface works and use it to achieve different tasks~\cite{Kelleher2005StencilsbasedTD, Grabler2009GeneratingPM, Li2013TutorialPlanAT}.   \dig offers the potential to automatically generate interactive tutorial walkthroughs because it manages all of the interface state and explicitly represents its correspondence to interactions in the UI.
Thus, given a start and end interface state---expressed as the states of the UI interactions and their corresponding set of bindings in the \dig grammar---we can automatically identify the sequence of user interactions necessary to go from start to end state, and use this sequence to generate an interactive, static, or video tutorial.

\begin{example}[Tutorial Generation]
Consider again the interface in \Cref{fig:intro} as the starting state and the following end state:
\begin{Verbatim}[fontsize=\footnotesize,samepage=true,commandchars=\\\{\}]
 SELECT year, payout1(*), ... FROM \red{evi}  WHERE dekad BETWEEN \red{1} AND \red{2}
\end{Verbatim}
To transition to the end state, we simply need to re-bind the choice variables \code{t} (using the dropdown) and  \code{s,e} (using the slider).    The order of interactions may be determined by e.g., a user cost model that estimates the amount of effort to perform different sequences.   
\end{example} 

More complex interface may contain data dependencies---where one choice variable $v_d$ may be a descendant of another $v_a$.  Given the \dig grammar, we can easily infer that the user must interact with $v_a$ before $v_d$.   

\subsection{Workload Generation}
Visualization benchmarks~\cite{Eichmann2018IDEBenchAB,battle2020database} are designed to help evaluate data processing systems that power interactive data interfaces by sequences of query workloads that simulate what an interface would produce during a user's analysis process.   However, existing benchmarks are limited in expressiveness---to SPJA query structures and parameterized filters.   Even simple transforms like changing the input relation (\Cref{fig:intro}) are not supported.   

In contrast, \dig can express arbitrary query structures, arbitrary transformations, and models a direct correspondence between user interactions and their query transformations.   As such, simply developing different user models---say, training a markov model or using a large language model to simulate an agent---can easily generate diverse query workloads and timings that reflect real data interfaces, queries, and user needs.

\section{Conclusions}

In this paper, we  propose \dig, a {\it Data Interface Grammar} that extends Parsing Expression Grammars (PEG) with annotations specific to data programs. 
\dig satisfies all three desired criteria: (C1) it can compactly express any set of queries useful for a task; (C2) it has a well-defined correspondence to interactive interfaces composed of charts, widgets,
and interactions; (C3) it is amenable to offline analysis.
We also demonstrate the compatibility with existing data engineering practices - DBT~\cite{dbt}. 
We further illustrate the potential benefits of this abstraction, such as automatic interface generation, 
automatic interface backend optimization, tutorial generation, and workload generation.
Addtionally, we describe how DIG simplifies interface creation via real-world examples.
\begin{acks}
Thanks to Miles Hong for helpful feedback.
This material is based upon work supported by NSF grants 1845638, 2008295, 2106197, 2103794; Amazon and Adobe.   Any opinions, findings, and conclusions or recommendations expressed in this material are those of the author(s) and do not necessarily reflect the views of the funders.
\end{acks}

\bibliographystyle{ACM-Reference-Format}
\bibliography{refs}
\end{document}